\documentclass[a4paper,12pt]{article}
\usepackage{epsfig}
\usepackage[dvips,usenames]{color}
\usepackage{graphicx}

\def\qslash{\rlap{\hspace{0.02cm}/}{q}}

\newlength{\dinwidth}
\newlength{\dinmargin}
\begin{document}
\title{Pair production of charged Higgs bosons in the Left-Right Twin Higgs model at the ILC and LHC}
\bigskip
\author{Yao-Bei Liu$^{1}$, Hong-Mei Han$^{1}$, Xue-Lei Wang$^{2}$\\
{\small 1: Henan Institute of Science and Technology, Xinxiang
453003, P.R.China}
\thanks{E-mail:hnxxlyb2000@sina.com}\\
 {\small 2: College of Physics and Information
Engineering,}\\
\small{Henan Normal University, Xinxiang  453007, P.R.China}
 }\maketitle
%\date{today}
\begin{abstract}
\indent Left-Right twin Higgs(LRTH) model predicts the existence of
a pair of charged Higgs $\phi^{\pm}$. In this paper, we study the
production of the charged Higgs bosons pair $\phi^{\pm}$ at the
International Linear Collider(ILC) and the $CERN$ Large Hadron
Collider(LHC). The numerical results show that the production rates
are at the level of several tens fb at the ILC, the process
$e^{+}e^{-}\rightarrow \phi^{+}\phi^{-}$ can produce the adequate
distinct multi-jet final states. We also discuss the charged Higgs
boson pair production via the process $q\bar{q}\rightarrow
\phi^{+}\phi^{-}$ at the LHC and estimate there production rates. We
find that, as long as the charged Higgs bosons are not too heavy,
they can be abundantly produced at the LHC. The possible signatures
of these new particles might be detected at the ILC and LHC
experiments.
\end{abstract}
PACS number(s): 12.60.Fr, 14.80.Mz; 14.65.Ha, 12.15.Lk
\newpage
\noindent{\bf I. Introduction}\\
 \indent The standard model(SM) provides an excellent
effective field theory description of almost all particle physics
experiments. But in the SM the Higgs boson mass suffers from an
instability under radiative corrections. The naturalness argument
suggests that the cutoff scale of the SM is not much above the
electroweak scale: New physics will appear around TeV energies.
Most extensions of the SM require the introduction of extended Higgs
sector to the theory. Generically, charged Higgs boson arises in the
extended Higgs sector, which does not exist in the SM. It implies
that the observation of a charged Higgs boson is a clear evidence
for existence of the new physics beyond the SM. Many alternative new
physics theories, such as supersymmetry, topcolor, and little Higgs,
predict the existence of new scalar or pseudo-scalor particles.
These new particles may have cross sections and branching fractions
that differ from those of the SM Higgs boson. Thus, studying the
production and decays of the new scalars
at high energy colliders will be of special interest.\\
  \indent Recently, the twin Higgs mechanism has
been proposed as a solution to the little hierarchy problem
\cite{ly-1,ly-2,ly-3}. The Higgs is a pseudo-Goldstone boson of a
spontaneously broken global symmetry. Gauge and Yukawa interactions
that explicitly break the global symmetry give mass to the Higgs.
Once a discrete symmetry is imposed, the leading quadratic divergent
term respects the global symmetry, thus does not contribute to the
Higgs model. The twin Higgs mechanism can be implemented in
left-right models with the discrete symmetry being identified with
left-right symmetry \cite{ly-2}. The left-right twin Higgs(LRTH)
model contains $U(4)_{1}\times U(4)_{2}$ global symmetry as well as
$SU(2)_{L}\times SU(2)_{R}\times U(1)_{B-L}$ gauge symmetry. The
left-right symmetry acts on only the two $SU(2)^{,}$s gauge
symmetry. The leading quadratically divergent SM gauge boson
contributions to the Higgs masses are canceled by the loop involving
the heavy gauge bosons. A vector top singlet pair is introduced to
generate an ${\cal O}(1)$ top Yukawa coupling. The quadratically
divergent SM top contributions to the Higgs potential are canceled
by the contributions from a heavy top partner. The LRTH model
predicted the existence of the new heavy particles, such as heavy
gauge boson, fermions, and scalars at or below the TeV scale, which
might generate characteristic signatures at the present and future
collider experiments \cite{Hock,dwj,liu}. \\
\indent The hunt for the Higgs and the elucidation of the mechanism
of symmetry
 breaking is one of the most important goals for present and future
 high energy collider experiments. Precision electroweak measurement
  data and direct searches suggest that the Higgs boson must be relative light and its mass should
 be roughly in the range of 114.4 GeV$\sim$208 GeV at $95\%$ CL \cite{Higgs}.
 While the discovery potential of the Higgs at the LHC has been established for a wide range of Higgs masses, only
 rough estimates on its properties will be possible, through measurements on the couplings of the Higgs
 to the fermions and gauge boson for example \cite{LHC}. The most precise measurements
 will be performed in the clean environment of the future high energy $e^{+}e^{-}$ linear
 collider, the International Linear Collider(ILC) with the center of mass(c.m.) energy $\sqrt{s}$=300
 GeV-1.5 TeV \cite{ILC} and the yearly luminosity 500 $fb^{-1}$. The
 running of the high energy and luminosity linear collider will
 open an unique window for us to understand the basic theory of
 particle physics. The aim of this paper is to investigate production of the charged Higgs bosons pair
 at the ILC and LHC and see whether the possible signatures of the LRTH model
 can be detected in the near future ILC and LHC experiments.\\
\indent This paper is organized as follows, The relevant couplings
to ordinary particles of the charged Higgs bosons $\phi^{\pm}$ in
the LRTH model are given in section II. In section III and section
IV we study the charged Higgs bosons pair productions at the ILC and
LHC, respectively. The numerical results and discussions are also
presented in theses sections. The conclusions are given in
section V. \\
\noindent{\bf II. The relative couplings}\\
 \indent The LRTH model is
based on the global $U(4)_{1}\times U(4)_{2}$ symmetry, with a
locally gauged subgroup $SU(2)_{L}\times SU(2)_{R}\times
U(1)_{B-L}$. Two Higgs fields, $H$ and $\hat{H}$, are introduced and
each transforms as $(4,1)$ and $(1,4)$ respectively under the global
symmetry. They are written as
\begin{eqnarray}
H=\left( \begin{array}{c} H_{L}\\
H_{R} \\
\end{array}  \right)\,,~~~~~~~~~~~~~~\hat{H}=\left( \begin{array}{c} \hat{H}_{L}\\
\hat{H}_{R} \\
\end{array}  \right)\,,
\end{eqnarray}
where $H_{L,R}$ and $\hat{H}_{L,R}$ are two component objects which
are charged under the $SU(2)_{L}\times SU(2)_{R}\times U(1)_{B-L}$
as
\begin{equation}
H_{L}~and~ \hat{H}_{L}: (2, 1, 1),~~~~~~~~H_{R}~ and~ \hat{H}_{R}:
(1, 2, 1).
\end{equation}
 The global
$U(4)_{1}(U(4)_{2})$ symmetry is spontaneously broken down to its
subgroup $U(3)_{1}(U(3)_{2})$ with non-zero vacuum expectation
values as $\langle H\rangle=(0,0,0,f)$ and $\langle
\hat{H}\rangle=(0,0,0,\hat{f})$. Each spontaneously symmetry
breaking results in seven Nambu-Goldstone bosons. Three of six
Goldstone bosons that are charged under $SU(2)_{R}$ are eaten by the
heavy gauge bosons, while leaves three physical Higgs: $\phi^{0}$
and $\phi^{\pm}$, The remaining Higgses are the SM Higgs doublet
$H_{L}$ and an extra Higgs doublet
$\hat{H}_{L}=(\hat{H}_{1}^{+},\hat{H}_{2}^{0})$ that only couples to
the gauge boson sector. A residue matter parity in the model renders
the neutral Higgs $\hat{H}_{2}^{0}$ stable, and it could be a good
dark matter candidate. A pair of vector-like quarks $q_{L}$ and
$q_{R}$ are introduced in order to give the top quark a mass of the
order of electroweak scale, which are singlets under
$SU(2)_{L}\times SU(2)_{R}$. The masse of the light SM-like top and
the heavy top are \cite{Hock}
\begin{eqnarray}
m_{t}^{2}&\simeq& y^{2}f^{2}\sin^{2}x-M^{2}\sin^{2}x\sim (y v/\sqrt{2})^{2}, \\
m_{T}^{2}&\simeq& y^{2}f^{2}+M^{2}-m_{t}^{2},
\end{eqnarray}
where $v=246GeV$ is the electroweak scale and $x=v/\sqrt{2}f$, the
mass parameter $M$ is essential to the mixing between the SM-like
top quark and the heavy top quark. The top Yukawa coupling can then
be determined by fitting the experimental value of the light top
quark mass. At the leading order of $1/f$, the mixing angles for
left-handed and right-handed fermions are \cite{Hock}
\begin{eqnarray}
S_{L}&\simeq& \frac{M}{m_{T}}\sin x,\\
S_{R}&\simeq& \frac{M}{m_{T}}(1+\sin^{2}x).
\end{eqnarray}
The value of $M$ is constrained by the requirement that the
branching ratio of $Z\rightarrow b\bar{b}$ remains consistent with
the experiments. We define the Weinberg angle in the $LRTH$ model
as\\
\begin{eqnarray}
s_{w}&=&\sin\theta_{w}=\frac{g'}{\sqrt{g^{2}+2g'^{2}}},\\
c_{w}&=&\cos\theta_{w}=\sqrt{\frac{g^{2}+g'^{2}}{g^{2}+2g'^{2}}}.
\end{eqnarray}
The unit of the electric charge is then given by
\begin{eqnarray}
e=gs_{w}=\frac{g g'}{g^{2}+2g'^{2}}.
\end{eqnarray}
\indent The $LRTH$ model introduces new charged Higgs bosons
$\phi^{\pm}$ in addition to the neutral Higgs $\phi^{0}$ and the SM
higgs boson $h$. For the neutral gauge bosons, we write the
couplings to fermions in the form
$i\gamma^{\mu}(g_{V}+g_{A}\gamma^{5})$. The couplings constants of
the heavy gauge bosons and the charged Higgs boson $\phi^{\pm}$ to
ordinary particles, which are related to our calculation, can be
written as \cite{Hock}
\begin{eqnarray}
g^{Z_{H}\bar{e}e}_{V}&=&\frac{e}{4s_{W}c_{W}\sqrt{1-2s^{2}_{W}}}(-1+4s^{2}_{W}),
\hspace{0.8cm}
g^{Z_{H}\bar{e}e}_{A}=-\frac{e}{4s_{W}c_{W}}\sqrt{1-2s^{2}_{W}},\\
g^{Z_{H}\bar{u}_{1,2}u_{1,2}}_{V}&=&\frac{e}{4s_{W}c_{W}\sqrt{1-2s^{2}_{W}}}(-1+\frac{8}{3}s^{2}_{W}),
\hspace{0.5cm}
g^{Z_{H}\bar{u}_{1,2}u_{1,2}}_{A}=-\frac{e}{4s_{W}c_{W}}\sqrt{1-2s^{2}_{W}},\\
g^{Z_{H}\bar{d}_{1,2}d_{1,2}}_{V}&=&\frac{e}{4s_{W}c_{W}\sqrt{1-2s^{2}_{W}}}(1-\frac{5}{3}s^{2}_{W}),
\hspace{0.7cm}
g^{Z_{H}\bar{d}_{1,2}d_{1,2}}_{A}=\frac{e}{4s_{W}c_{W}^{3}\sqrt{1-2s^{2}_{W}}},\\
g^{Z_{H}\bar{b}b}_{V}&=&\frac{e}{4s_{W}c_{W}\sqrt{1-2s^{2}_{W}}}(1-\frac{4}{3}s^{2}_{W}),
\hspace{0.7cm}
g^{Z_{H}\bar{b}b}_{A}=\frac{e}{4s_{W}c_{W}}\sqrt{1-2s^{2}_{W}},\\
g^{\phi^{-}\phi^{+}\gamma}&=&-e(p_{1}-p_{2})_{\mu}, \hspace{3.3cm}
g^{\phi^{-}\phi^{+}Z}=\frac{es_{W}}{c_{W}}(p_{1}-p_{2})_{\mu},\\
g^{\phi^{-}\phi^{+}Z_{H}}&=&-\frac{e(1-3s^{2}_{W})}{2s_{W}c_{W}\sqrt{1-2s^{2}_{W}}}(p_{1}-p_{2})_{\mu},
\hspace{0.7cm} g^{\phi^{+}
\bar{t}b}=-i(S_{R}m_{b}P_{L}-yS_{L}fP_{R})/f,
\end{eqnarray}
where $P_{L(R)}=(1\mp \gamma_{5})/2$ is the left(right)-handed
projection operator. $s_{W}(c_{W})$
represents the sine(cosine)of the Weinberg angle, $p_{1}$ and $p_{2}$ refer to the incoming momentum of the first and second particle, respectively. \\
\noindent{\bf III Production of the charged Higgs bosons pair at
ILC}\\
 \indent From above discussions, we can see that the charged
Higgs bosons pair $\phi^{+}\phi^{-}$ can be produced in $e^{+}e^{-}$
annihilation via virtual photon or
 neutral gauge boson exchanging. The Feynman diagram of the process $e^{+}e^{-}\rightarrow \phi^{+}\phi^{-}$ is shown in Fig.1.\\
 \begin{figure}[ht]
\begin{center}
\epsfig{file=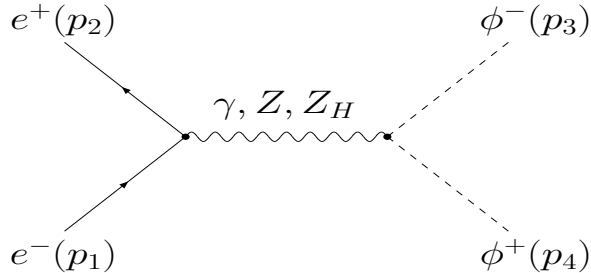,width=450pt,height=500pt} \vspace{-13cm}
\caption{\small The Feynman diagrams of the process
$e^{+}e^{-}\rightarrow \phi^{+}\phi^{-}$.} \label{fig1}
\end{center}
\end{figure}
\indent At the leading order, the production amplitude of the
process can be written as:
\begin{equation}
 \mathcal{M}_{1}=\mathcal{M}_{\gamma}+\mathcal{M}_{Z}+\mathcal{M}_{Z_{H}},
 \end{equation}
 with
 \begin{eqnarray*}
 \mathcal{M}_{\gamma}&=&e^{2}\overline{v}_{e}(p_{2})\gamma^{\mu}u_{e}(p_{1})G^{\mu\nu}(p_{1}+p_{2},0)(p_{3}-p_{4})_{\nu},\\
 \mathcal{M}_{Z}&=&\frac{es_{W}}{c_{W}}\overline{v}_{e}(p_{2})\gamma^{\mu}(g^{Z\bar{e}e}_{V}+^{Z\bar{e}e}_{A}\gamma^{5})u_{e}(p_{1})
G^{\mu\nu}(p_{1}+p_{2},M_{Z})(p_{3}-p_{4})_{\nu},\\
\mathcal{M}_{Z_{H}}&=&\frac{e(1-3s^{2}_{W})}{2s_{W}c_{W}\sqrt{1-2s^{2}_{W}}}\overline{v}_{e}(p_{2})\gamma^{\mu}(g^{Z_{H}\bar{e}e}_{V}+^{Z_{H}\bar{e}e}_{A}\gamma^{5})u_{e}(p_{1})
G^{\mu\nu}(p_{1}+p_{2},M_{Z_{H}})(p_{3}-p_{4})_{\nu}.
 \end{eqnarray*}
Here, $G^{\mu\nu}(p,M)=\frac{-ig^{\mu\nu}}{p^{2}-M^{2}}$ is the
propagator of the particle. With above production amplitudes, we can
obtain the production cross section directly. In the calculation of
the cross section, instead of calculating the square of the
amplitudes analytically, we calculate the amplitudes numerically by
using the method of the references\cite{HZ} which can greatly
simplify our calculation. \\
\indent In the numerical calculation, we take the input parameters
as $G_{F}=1.166\times10^{-5}GeV^{-2}$, $M_{Z}=91.187$ GeV and
$s_{W}^{2}=0.2315$ \cite{data}. The electromagnetic fine-structure
constant $\alpha_{e}$ at a certain energy scale is calculated from
the simple QED one-loop evolution with the boundary value
$\alpha_{e}=1/137.04$. Except for these $SM$ input parameters, the
production cross sections is dependent on the symmetry breaking $f$,
the mixing parameter M and the mass of the charged Higgs boson
$\phi^{\pm}$. In our analysis, we take $M$ to be small and pick a
typical value of $M=150 GeV$. The symmetry breaking scales $f$ is
allowed in the range of $500\sim 1500 GeV$ \cite{Hock}.
 As numerical estimation, we will assume that $M_{\phi^{\pm}}$ is in the range of $150 \sim 450 GeV$. \\
\begin{figure}[ht]
\begin{center}
\scalebox{0.9}{\epsfig{file=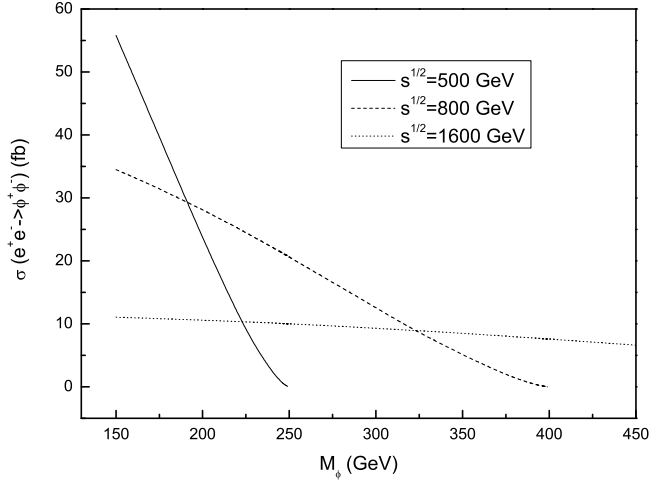}}\\
\end{center}
\caption{\small The cross section of $e^{+}e^{-}\rightarrow
\phi^{+}\phi^{-}$ as a function of charged Higgs bosons mass
$M_{\phi^{\pm}}$ for $f=1 TeV$, $M=150 GeV$ and various
center-of-mass $\sqrt{s}$.}
\end{figure}
\indent In Fig.2, we plot the cross section of
$e^{+}e^{-}\rightarrow \phi^{+}\phi^{-}$ are plotted as a function
of the mass parameter $M_{\phi^{\pm}}$ for $M_{Z_{H}}=2.5 TeV$ and
three values of the the center of mass energy. The plots show that
the cross section decrease with $M_{\phi^{\pm}}$ due to the phase
space suppression. For $\sqrt{s}=500 GeV$, the cross section falls
sharply to a very small rate with $M_{\phi^{\pm}}$ increasing. So,
the energy 500 GeV is not suitable to search for the heavy charged
Higgs bosons pair. The cross section is not sensitive to
$M_{\phi^{\pm}}$ when $\sqrt{s}=1600 GeV$. The change of the cross
section with $\sqrt{s}$ is not monotonous because the influence of
$\sqrt{s}$ on the phase space and Z-propagator is inverse. In the
most case, the production rate is at the order of tens fb. For the
light charged scalars, the production rate can be near 30fb in the
case of $\sqrt{s}=800 GeV$. With yearly expected luminosity about
$\pounds=500fb^{-1}$, then there will be $10^{2}\sim 10^{4}$
events to be generated each year.\\
\begin{figure}[ht]
\begin{center}
\scalebox{0.9}{\epsfig{file=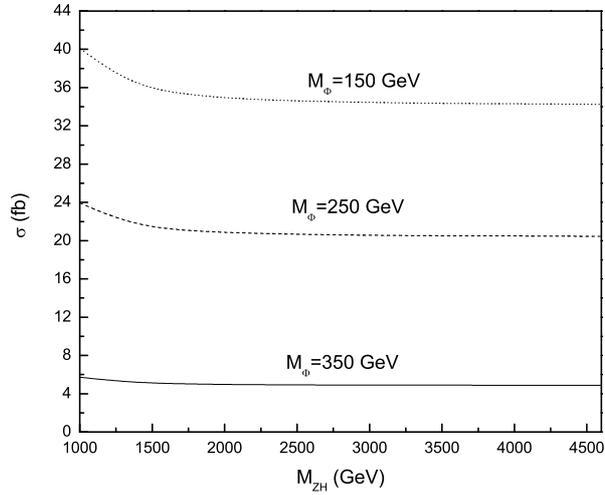}}\\
\end{center}
\caption{\small The cross section $\sigma(s)$ of
$e^{+}e^{-}\rightarrow \phi^{+}\phi^{-}$ as a function of the heavy
gauge boson mass $M_{Z_{H}}$ for $\sqrt{s}$=800 GeV and different
values of $M_{\phi^{\pm}}$.}
\end{figure}
\indent To see the influence of the new heavy gauge boson mass
$M_{Z_{H}}$ on the cross section, in Fig.3 we plot $\sigma(s)$ as a
function of $M_{Z_{H}}$ for $\sqrt{s}$=800 GeV and three values of
$M_{\phi^{\pm}}$=150, 250, 350 GeV, respectively. From Fig.3, one
can see that the cross section decreases slowly with $M_{Z_{H}}$
increasing and is more sensitive to the charged Higgs bosons mass.
This is because the production cross section are mainly aroused by
the exchange of $\gamma$ and $Z$ boson. In general, the cross
section is at the order of tens fb. This abundant production allows
to enforce tight requirements on the event pre-selection and
the mass reconstruction. \\
\indent It has been shown that the charged Higgs $\phi^{\pm}$
dominantly decay into $tb$ for larger value of the mixing parameter
between the SM-like top quark and the heavy top quark \cite{Hock}.
In the case of $\phi^{+}\rightarrow t\bar{b}$, the signals of the
charged Higgs bosons pair production is $t\bar{t}b\bar{b}$. The
cross section of the irreducible $t\bar{t}b\bar{b}$ background has
been estimated using the Comphep program \cite{comphep} at 0.8 TeV
and found to be 5.5 fb. In order to efficiently distinguish the
signals from the underlying backgrounds and to measure the charged
scalar mass, it is important to obtain a clean charged scalar signal
in the mass distribution of the multi-jet final states. To identify
the production mode $t\bar{t}b\bar{b}$, we insist on 8 jets or 1
lepton plus 6 jets(in
 particular, fewer than 10 visible lepton/jets so as to discriminate from the 4t final states) and possibly require that one
  W and the associated top quark be reconstructed. In particular, since final states contain at least four b jets, in order
  to eliminate any residual QCD background, we need one or two b-tags without incurring significant
  penalty. Such b-tagging should have efficiency of $60\%$ or
  better. The mistagging of b-quark and s-quark will make the $e^{+}e^{-}\rightarrow W^{+}W^{-}$
  become important which significantly enhance the background. So,
the efficient b tagging and mass reconstruction
  of the charged Higgs bosons is very necessary to reduce the background \cite{marco}.\\
\indent It is known that many new physics model predict similar
heavy charged scalars, such as $\Pi^{\pm}$ in the topcolor-assisted
technicolor model(TC2) and $H^{\pm}$ in the two-Higgs doublet
model(2HDM). To distinguish the scalars in the LRTH model from the
charged top-pions in the TC2 model and the Higgs in the 2HDM, we
should compare the cross section of $e^{+}e^{-}\rightarrow
\phi^{+}\phi^{-}$ with those of similar process
$e^{+}e^{-}\rightarrow \Pi^{+}\Pi^{-}$ \cite{wang} in TC2 model and
process
 $e^{+}e^{-}\rightarrow H^{+}H^{-}$ \cite{2hdm} in the 2HDM. The cross section sections values of such three process are
 not significantly different for the same parameters. For example, $\sigma(e^{+}e^{-}\rightarrow \Pi^{+}\Pi^{-})=21.51, 15.09 fb$ with
 $\sqrt{s}$=800, 1600 GeV and $M_{\Pi}$=300 GeV, $\sigma(e^{+}e^{-}\rightarrow H^{+}H^{-})=10.02, 8.1 fb$ with
 $\sqrt{s}$=800, 1600 GeV and $M_{H}$=300 GeV, $\sigma(e^{+}e^{-}\rightarrow \phi^{+}\phi^{-})=12.54, 9.26 fb$ with
 $\sqrt{s}$=800, 1600 GeV and $M_{\phi}$=300 GeV. So we
 should distinguish them depending on their different feature of decay modes and pole
 structure. The $t\bar{b}$ is the main decay mode for the charged
 scalars. However, we should probe charged top-pions via the
 flavor-changing decay mode $\Pi^{+}\rightarrow c\bar{b}$ to obtain
 the identified signals. $\tau\nu_{\tau}$ can also provide the
 identified signals of charged Higgs from the 2HDM which is not
 exist for the charged top-pions and $\phi^{\pm}$.\\
\noindent{\bf IV. Production of the charged Higgs bosons pair at
LHC}\\
 \indent The Large Hadron Collider(LHC) at CERN has a good potential for discovery of a charged Higgs boson.
 At the LHC, the charged Higgs bosons also can be produced in pair production mode.
 There are two important $\phi^{+}\phi^{-}$ production channels: (i) $q\bar{q}\rightarrow
\phi^{+}\phi^{-}$, where $(q=u$, $d$, $c$, $s$, $b)$(via Drell-Yan
process, where a photon and a Z-boson are exchanged in the
$s$-channel and the top quark in the t-channel.)\cite{eeich} (ii)
the loop-induced gluon fusion process $gg\rightarrow
\phi^{+}\phi^{-}$ \cite{aab}. The Feynman diagrams of these
processes are shown in Fig.4. In the case of $q=b$, there are
additional Feynman diagrams involving $\phi^{0}$ and $H$ in the
 Fig.4(a), the neutral Higgs bosons $\phi^{0}$ and $H$ exchange
in the $s$-channel can also contribute to the pair production
process. However, these contributions are very smaller than those of
other tree-level processes because of either the small Yukawa
couplings, small patron distribution functions or both combined. On
the other hand, the contributions from the Fig.4.(c-e) are also very
smaller than those of the tree-level processes. This is because the
Yukawa couplings depends sensitively on the parameter $M$ and $f$.
For small $f$, the values of the parameter $M$ are very small
\cite{dwj}. Once $M$ is very small or in the limit that $M$=0,
certain couplings go to zero. Although the gluon fusion get an
enhancement due to larger patron distribution functions, the
contributions of gluon fusion process is suppressed by the order of
$(M/f)^{4}$.  Thus, we will
ignore these processes in the following estimation.\\
 \begin{figure}[t]
\begin{center}
\epsfig{file=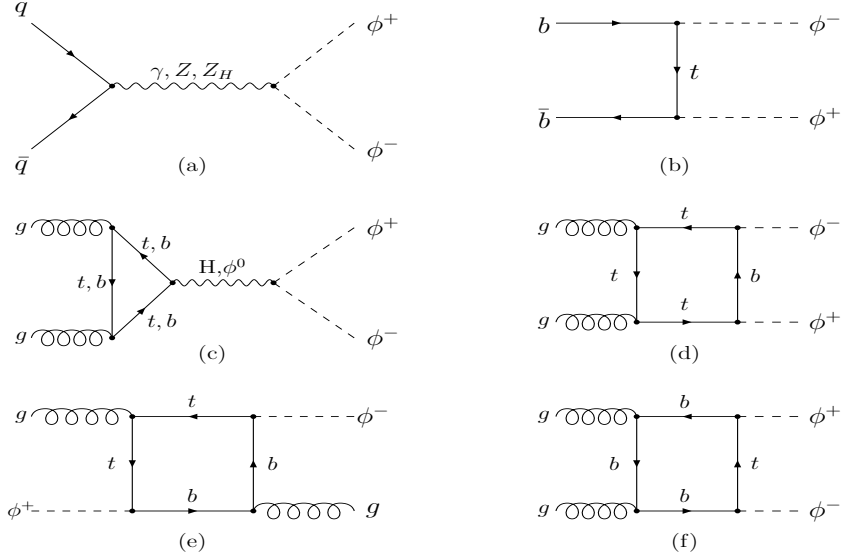,width=450pt,height=500pt} \vspace{-8.5cm}
\caption{\small The tree-level Feynman diagrams for the process
$q\bar{q}\rightarrow \phi^{+}\phi^{-}$(Fig.(a-b)) and the one-loop
Feynman diagrams for $gg \rightarrow \phi^{+}\phi^{-}$(Fig.(c-f)) in
the LRTH model.} \label{fig1}
\end{center}
\end{figure}
\indent Using the relevant Feynman rules, we can write the invariant
amplitude for the parton process $q(p_{1})\bar{q}(p_{2})\rightarrow
\phi^{+}(p_{3})\phi^{-}(p_{4})$ as:
\begin{eqnarray}
 \mathcal{M}_{2}= \mathcal{M}_{21}+ \mathcal{M}_{22}
\end{eqnarray}
\indent For the process $b\bar{b}\rightarrow \phi^{+}\phi^{-}$, the
invariant amplitude comes from the Fig.4(a) and (b):
\begin{eqnarray}
 \mathcal{M}_{21}&=&e\bar{v}(p_{2})Q\gamma_{\nu}u(p_{1})g^{\mu\nu}(p_{4}-p_{3})_{\mu}\nonumber\\
& &+
\frac{es_{W}}{c_{W}}\bar{v}(p_{2})\gamma_{\nu}(g_{V}^{Zb\bar{b}}+g_{A}^{Zb\bar{b}}\gamma_{5})u(p_{1})\frac{g^{\mu\nu}}{\hat{s}-m_{Z}^{2}}(p_{4}-p_{3})_{\mu}\nonumber\\
& &+
\frac{e(1-3s_{W}^{2})}{2c_{W}s_{W}\sqrt{1-2s_{w}^{2}}}\bar{v}(p_{2})\gamma_{\nu}(g_{V}^{Z_{H}b\bar{b}}+g_{A}^{Z_{H}b\bar{b}}\gamma_{5})u(p_{1})\frac{g^{\mu\nu}}{\hat{s}-m_{Z_{H}}^{2}}(p_{4}-p_{3})_{\mu}\nonumber\\
&
&+\frac{1}{f^{2}}\bar{v}(p_{2})(S_{R}m_{b}P_{L}-yS_{L}fP_{R})\frac{\qslash+m_{t}}{\hat{t}-m_{t}^{2}}(S_{R}m_{b}P_{R}-yS_{L}fP_{L})u(p_{1})
\end{eqnarray}
\indent For $u$, $c$, $d$, $c$ and $s$ quarks, we only consider the
contributions of the s-channel process to the scattering amplitude,
which can be written as:
\begin{eqnarray}
 \mathcal{M}_{22}&=&e\bar{v}(p_{2})Q\gamma_{\nu}u(p_{1})g^{\mu\nu}(p_{4}-p_{3})_{\mu}\nonumber\\
& &+
\frac{es_{W}}{c_{W}}\bar{v}(p_{2})\gamma_{\nu}(g_{V}^{Zq\bar{q}}+g_{A}^{Zq\bar{q}}\gamma_{5})u(p_{1})\frac{g^{\mu\nu}}{\hat{s}-m_{Z}^{2}}(p_{4}-p_{3})_{\mu}\nonumber\\
& &+
\frac{e(1-3s_{W}^{2})}{2c_{W}s_{W}\sqrt{1-2s_{w}^{2}}}\bar{v}(p_{2})\gamma_{\nu}(g_{V}^{Z_{H}q\bar{q}}+g_{A}^{Z_{H}q\bar{q}}\gamma_{5})u(p_{1})\frac{g^{\mu\nu}}{\hat{s}-m_{Z_{H}}^{2}}(p_{4}-p_{3})_{\mu}
\end{eqnarray}
with $Q=2e/3$ (for $q=u, c$) and $Q=-e/3$ (for $q=d, s, b$). Where
$\hat{s}=(p_{1}+p_{2})^{2}$ and $\hat{t}=(p_{1}-p_{3})^{2}$ are the
usual Mandelstam variables. We have neglected the light quark masses
in our calculations except the bottom quark. From above equations,
we can see that the production cross of the t-channel process is
suppressed by the order of $(M/f)^{4}$. Furthermore, the cross
section via the new heavy gauge boson $Z_{H}$ exchange is suppressed
by an order of magnitude compared to that for the s-channel Z
exchange and photon exchange. Therefore, the main production
processes are the usual Drell-Yan processes through the s-channel Z
exchange and
photon exchange. \\
\begin{figure}[t]
\begin{center}
\scalebox{0.9}{\epsfig{file=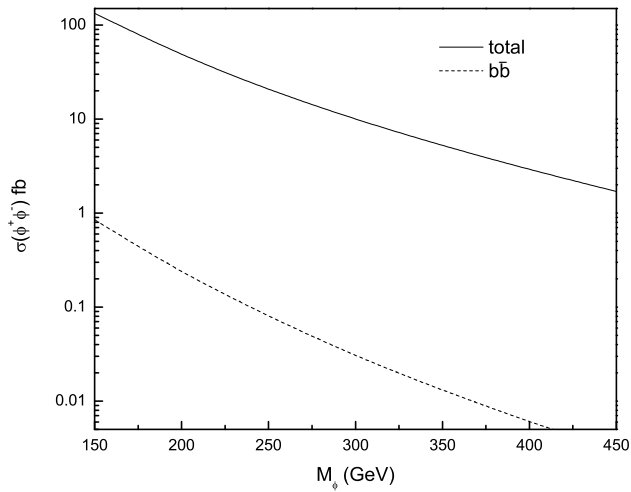}}\\
\end{center}
\caption{\small The cross section of the process
$q\bar{q}\rightarrow \phi^{+}\phi^{-}$ as a function of charged
Higgs bosons mass $M_{\phi^{\pm}}$ for $f=1000 GeV$ and $M=150
GeV$.}
\end{figure}
\indent To get the numerical results, we take the input parameters
as $m_{t}=172.7GeV$ \cite{topmass} and used the CTEQ6L patron
distribution functions \cite{fun} and two-loop running coupling
constant $\alpha_{s}(m_{Z})=0.118$. There are two free parameters
$f$ and the value of the mixing parameter $M$. In this paper, we
will take the typical values of $f$=1000 GeV and $M$=150 GeV. Our
numerical results are shown in Fig.(5), in which we plot the
production cross section $\sigma(\phi^{+}\phi^{-})$ for the process
$p\bar{p}\rightarrow \phi^{+}\phi^{-}+X$ at the LHC with
$\sqrt{s}=14 TeV$ as a function of charged Higgs bosons mass
$M_{\phi^{\pm}}$ for $f=1000 GeV$. To comparison, we use the solid
line and dashed line to represent the contributions of the process
$q\bar{q}\rightarrow \phi^{+}\phi^{-}$$(q=u$, $d$, $c$, and $s$) and
the process $b\bar{b}\rightarrow \phi^{+}\phi^{-}$, respectively.
From Fig.5 one can see that the production cross section of the
charged Higgs bosons $\phi^{+}\phi^{-}$ mainly comes from the usual
Drell-Yan processes $q\bar{q}\rightarrow \phi^{+}\phi^{-}$$(q=u$,
$d$, $c$, and $s$) through the s-channel gauge bosons exchange and
photon exchange. The total production cross section
$\sigma(\phi^{+}\phi^{-})$ is in the range of $134.5 fb\sim 1.7 fb$
for $150 GeV \leq M_{\phi^{\pm}}\leq 450 GeV$. The charged top-pions
pair in the Higgsless-top-Higgs(HTH) model and charged Higgs bosons
in the minimal supersymmetry standard moedl(MSSM) pair production at
the LHC have been calculated to leading and next-to-leading order
\cite{yue,next}. They have shown that the total cross section for
the charged top-pions and Higgs bosons pair production processes is
smaller than $10fb$ in most of the parameter spaces. Thus, we
expected that the charged Higgs bosons $\phi^{+}\phi^{-}$ predicted
by the LRTH model can be more easy detected at the LHC via this
process than those for the charged top-pions $\pi^{\pm}$ in HTH
model and charged Higgs bosons $H^{\pm}$ in the MSSM.\\
\noindent{\bf V. Conclusions}\\
 \indent The SM predicts the existence of a neutral Higgs boson,
while many popular models beyond the SM predict the existence of the
neutral or charged scale particles. These new particles might
produce the observable signatures in the current or future high
energy experiments, which is different from that for the SM Higgs
boson. Any visible signal from the new scalar particles will be
evidence of new physics beyond the SM. Thus, studying the new scalar
particle production is very interesting at the ILC and LHC.\\
\indent The twin Higgs mechanism provides an alternative method to
solve the little hierarchy problem. The LRTH model is a concrete
realization of the twin Higgs mechanism. The cancelations of
divergences occurs by alignment of vacua and existence of several
new particles. The new particles in the LRTH model are heavy top
quark, new gauge
 bosons, and new Higgs bosons, which might produce characteristic
 signatures at the ILC and LHC experiments.\\
 \indent In this paper, we discuss the pair production of the charged Higgs bosons $\phi^{+}\phi^{-}$ predicted by the LRTH model at the ILC
and the LHC via suitable mechanisms. We can obtain the following
conclusions: (i) For the production of $\phi^{+}\phi^{-}$ at the
ILC, we found that the production
 rate is at the level of several tens fb in a large of the parameter space. The efficient b tagging and mass reconstruction
  of the charged Higgs bosons is needed in order to reduce the background. We concluded that
 the charged scalars $\phi^{\pm}$ predicted by the LRTH model should
 be experimentally observable via the process $e^{+}e^{-}\rightarrow
 \phi^{+}\phi^{-}$ at the ILC. (ii) For the production of
$\phi^{+}\phi^{-}$ at the LHC, we found that the total production
cross section $\sigma(\phi^{+}\phi^{-})$ is in the range of $134.5
fb\sim 1.7 fb$ for $150 GeV \leq M_{\phi^{\pm}}\leq 450 GeV$, which
might be larger than those for the charged top-pions $\pi^{\pm}$ in
HTH model and charged Higgs bosons $H^{\pm}$ in the MSSM. The main
production processes are the usual Drell-Yan processes through the
s-channel Z exchange and photon exchange. In conclusion, as long as
the charged Higgs bosons is not too heavy, we conclude that the pair
production of the charged Higgs bosons will be a good test for the
LRTH model at future ILC and LHC experiments.

\vspace{.5cm} \noindent{\bf Acknowledgments}

This work is supported in part by the National Natural Science
Foundation of China(Grant No.10775039 and 10575029) and a grant from
Henan Institute of Science and Technology(06040).

\end{document}